\documentclass[reprint, superscriptaddress, secnumarabic, amssymb, nobibnotes, aps, prl]{revtex4-1}

\usepackage{graphicx}
\usepackage{epstopdf}
\usepackage[T1]{fontenc}
\usepackage{amsbsy}
\usepackage{gensymb}
\usepackage[T1]{fontenc}
\usepackage{amsmath}
\usepackage{amssymb}
\usepackage{bbm}
\usepackage{braket}
\usepackage{xcolor}
\allowdisplaybreaks
\usepackage{graphicx}
\usepackage{natbib}
\usepackage[colorlinks=true]{hyperref}  
\hypersetup{
    bookmarks=true,      
    unicode=false,     
    pdftoolbar=true,        
    pdfmenubar=true,        
    pdffitwindow=false,     
    pdfstartview={FitH},    
    pdftitle={Superconductivity in LixTaS2},    
    pdfauthor={Tarushi Agarwal},      
    pdfsubject={},   
    pdfcreator={},   
    pdfproducer={}, 
    pdfkeywords={} {} {}, 
    pdfnewwindow=true,   
    colorlinks=true,       
    linkcolor=blue,          
    citecolor=blue,        
    filecolor=magenta,      
    urlcolor=blue           
} 

\usepackage[normalem]{ulem}

\newcommand{\equref}[1]{Eq.~(\ref{#1})}

\newcommand{\figref}[1]{Fig.~\ref{#1}}

\begin{document}

\title{\textrm{Quasi-two-dimensional anisotropic superconductivity in Li intercalated 2H-TaS$_2$}}
\author{T. Agarwal}
\affiliation{Department of Physics, Indian Institute of Science Education and Research Bhopal, Bhopal, 462066, India}
\author{C. Patra}
\affiliation{Department of Physics, Indian Institute of Science Education and Research Bhopal, Bhopal, 462066, India}
\author{A. Kataria}
\affiliation{Department of Physics, Indian Institute of Science Education and Research Bhopal, Bhopal, 462066, India}
\author{Rajeshwari R. Chaudhari}
\affiliation{Department of Physics, Indian Institute of Science Education and Research Bhopal, Bhopal, 462066, India}
\author{R. P. Singh}
\email[]{rpsingh@iiserb.ac.in}
\affiliation{Department of Physics, Indian Institute of Science Education and Research Bhopal, Bhopal, 462066, India}

\begin{abstract}
Two-dimensional (2D) superconductivity in artificial interfaces and atomic-thin layers has gained attention for its exotic quantum phenomena and practical applications. Although bulk van der Waals layered materials have been explored for 2D superconductivity, most of these compounds do not exhibit remarkable properties despite exhibiting 2D characteristics. Here we report a comprehensive analysis of single crystals of Li intercalated 2H-TaS$_2$ superconductor, suggesting weakly coupled anisotropic superconductivity. Angle-dependent magnetotransport and the Berezinskii-Kosterlitz-Thouless (BKT) transition confirm quasi-2D superconductivity in 2H-Li$_x$TaS$_2$.
\end{abstract}
\maketitle
\section{INTRODUCTION}
Recent studies on superconductivity in reduced dimension have drawn increasing attention to exploring new phenomena such as quantum metallic ground state, quantum Griffith singularity, and Ising superconductivity \cite{2Dsuperconductors,qmat_gs,qmat_gs2,Ising,Ising_mos2}. Superconductivity in two-dimensional materials has been observed in materials where their thickness is less than the superconducting coherence length ($d$ < $\xi$), such as in thin films and hetero-structured interfaces \cite{2d_interfaces,LixNbO2,2d_at_interface}. However, achieving superconductivity in materials with an atomic limit thickness, where it is susceptible to disorders, poses a significant challenge. Anisotropic layered materials, such as MoS$_2$, NbSe$_2$, and ZrNCl, offer a promising platform to study 2D superconductivity and have exhibited unconventional properties \cite{QPT_vdw,ZrNCl}. Recently, clean limit 2D superconductivity has been discovered in bulk Ba$_6$Nb$_{11}$S$_{28}$ due to the presence of sub-layers NbS$_2$ and Ba$_3$NbS$_5$ \cite{Ba6Nb11S28}.

Layered transition metal dichalcogenides (TMDs) have attracted particular interest due to their broad range of properties, including dimension-dependent band structure, polymorphic phase transitions, and significant electron-electron and electron-phonon interactions \cite{polymorphic,2Dmaterials,TmdCs}. These systems exhibit novel phenomena, such as the coexistence of charge density wave (CDW) and superconductivity, metal-insulator transition, spin-valley coupling, and quantum spin hall effect, due to their 2D nature and the strong spin-orbit coupling effect caused by heavy metal atoms \cite{cdw-SC,cdw_NbSe2,QSH}. Moreover, TMDs in the few-layer limit exhibit extremely high in-plane upper critical field ($H_{c2}$) values that exceed the Pauli limiting field several times, which can be attributed to their inherent strong spin-orbit coupling and inversion symmetry breaking, leading to exotic forms of superconductivity \cite{Ising_mos2,2D-TaS2}. However, the fabrication processes for low-dimensional materials are typically complicated, and this may hinder their characterization through various techniques, which is essential for a complete understanding of the phenomena. Intercalation or chemical doping of these materials represents an alternate way to achieve 2D superconductivity by modifying interlayer coupling and electronic properties. Reports have shown that quasi-2D superconductivity can be achieved by adjusting the interlayer coupling in materials such as AuTe$_2$Se$_{4/3}$, organic ion intercalated SnSe$_2$, and Pb$_{1/3}$TaS$_2$ \cite{AuTe2Se,SnSe2_int,Pb0.33TaS2}.

TaS$_2$ is an important member of the TMD family, with various polytypes determined by the arrangement of Ta and S atoms, exhibiting unique and fascinating properties. The 2H-TaS$_2$ family displays chiral charge ordering \cite{Guillam_n_2011} while substituting S atoms with Se atoms in 2H-TaSeS leads to 2D multigap superconductivity that breaks the Pauli limit \cite{TaSSe}. Furthermore, intercalated 2H-TaS$2$ with organic molecules like (Py)${0.5}$TaS$_2$, where Py represents pyridine, has shown a dimensional crossover from 3D to 2D \cite{TaS2_int}.

In this letter, we have reported single crystal growth and superconducting properties of Li intercalated 2H-TaS$_2$ compound using transport, magnetisation, and specific heat measurements. It confirms weakly coupled anisotropic superconductivity having bulk superconducting transition at $T_c$ = 3.3(1) K. Detailed specific heat measurement identifies Li$_{x}$TaS$_{2}$ as a weakly coupled superconductor. Angle-dependent magnetotransport and Berezinskii-Kosterlitz-Thouless (BKT) transition ($T_\text{BKT}$ = 2.9 K) confirm the quasi-two-dimensional (2D) superconductivity in 2H-Li$_x$TaS$_2$.

\section{EXPERIMENTAL DETAILS}
\begin{figure*}[!t]
\centering
\includegraphics[width=1.0\textwidth]{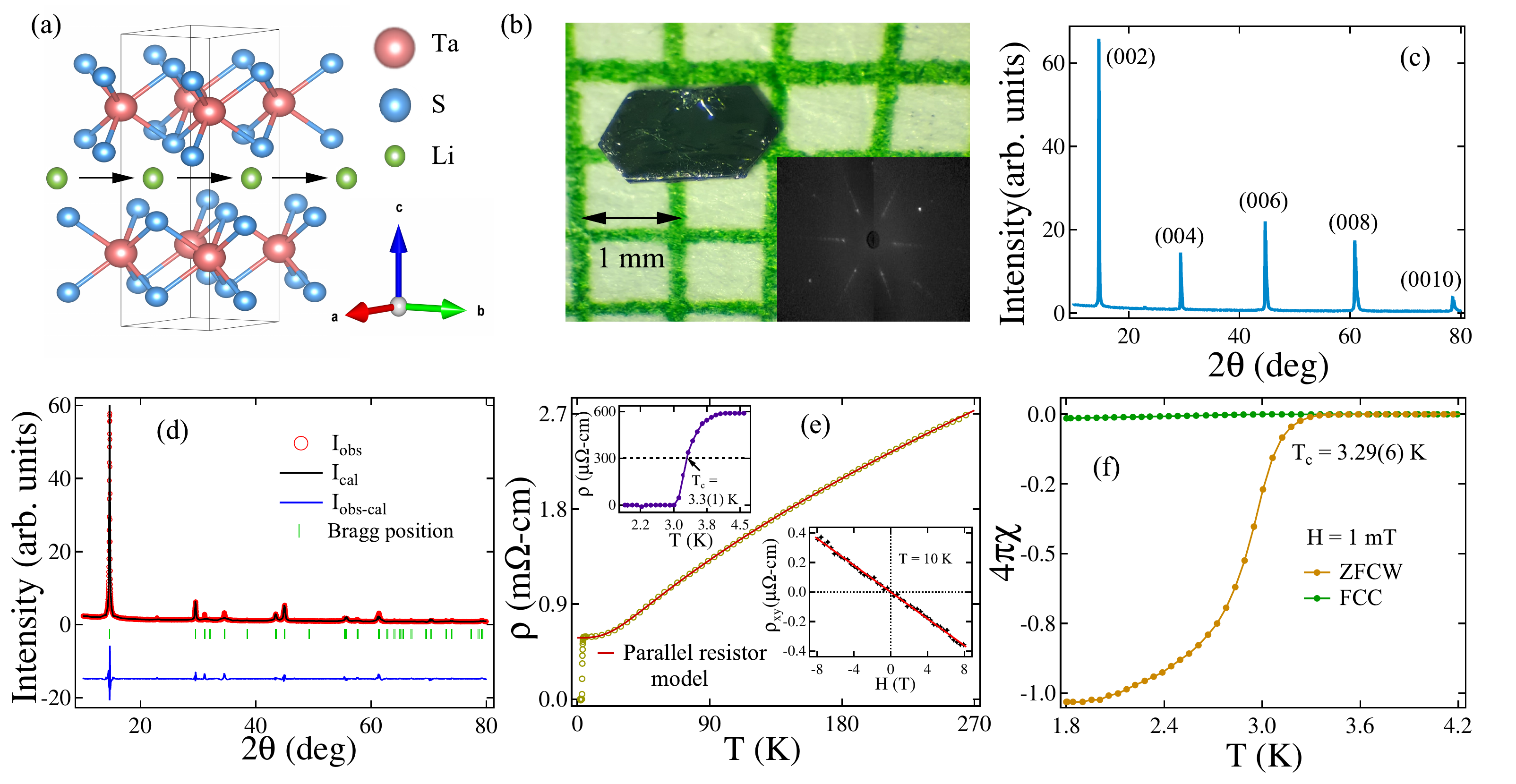}
\caption{\label{Fig1}(a) The crystal structure of Li intercalated 2H-TaS$_2$. (b) The single crystal image of Li intercalated 2H-TaS$_2$ and the inset shows the Laue diffraction pattern of the crystal. (c) shows single crystal XRD with labelled $[00l]$ reflections indicating the $c$-axis as the growth direction. (d) Powder XRD of well-crushed crystals refined with a 2H-TaS$_2$ hexagonal structure. (e) Temperature-dependent electrical resistivity at zero field fitted with the parallel resistor model. The inset (upside) shows zero drop transition of resistivity in the superconducting region observed in the $H \parallel c$ direction. The lower inset shows the hall resistivity at $T$ = 10 K with a negative slope, indicating n-type charge carriers in Li$_x$TaS$_2$. (f) Temperature dependence of magnetic susceptibility in both ZFCW and FCC modes measured at $H$ = 1 mT.}
\end{figure*}

Single crystals of the Li$_{x}$TaS$_{2}$ compound were synthesized using the chemical vapour transport (CVT) method. The starting materials, Ta powder (99.9$\%$) and S powder (99.5$\%$), were used in the required ratio with LiCl as a transporting agent. Li$_{x}$TaS$_2$ was prepared with $x$ = 0.3 of Li content. The powder was sealed in a quartz tube after the tube was evacuated to $\sim$ 10$^{-6}$ mbar. The sealed tube was placed in a two-zone tubular furnace and heated at a temperature gradient of 1083 K-1053 K for 10 days and then slowly cooled to room temperature. Shiny millimetre-sized crystals were obtained in the cold zone.

Powder X-ray diffraction (XRD) measurements were performed on a single crystalline sample and crushed crystals using an X’pert PANalytical diffractometer with monochromatic Cu-K$_\alpha$ radiation ($\lambda$ = 1.5406 $\text{\AA}$) to determine the structure and phase purity of the sample. The Laue diffraction pattern was recorded using a Photonic–Science Laue camera system. A Quantum Design magnetic measurement system (MPMS3) was used to measure magnetization. Transport and specific heat measurements were also conducted using a Quantum Design physical property measurement system (PPMS). A Quantum Design horizontal rotator insert was used for angle-dependent magnetotransport measurements, and all transport data were collected using the standard four-probe method.

\section{RESULTS AND DISCUSSION}

\subsection{Sample characterization}

\figref{Fig1}(a) shows the crystal structure of 2H-TaS$_2$ where Li is intercalated within the layers. In the 2H phase, the Ta atom is coordinated in a trigonal prismatic pattern with chalcogen S atoms. \figref{Fig1}(b) shows the crystal image obtained by the CVT method with a Laue pattern in the inset. The single-crystal XRD pattern of Li$_{x}$TaS$_{2}$ is shown in \figref{Fig1}(c), reflecting crystal growth in the $c$-direction with $[00l]$ peaks. Powder XRD was performed on well-crushed crystals. The observed, calculated, and difference Rietveld profiles for powdered Li$_{x}$TaS$_2$ have been shown in \figref{Fig1}(d) with Bragg positions. To deduce the lattice parameters, we performed XRD refinement using Fullprof software. Li$_{x}$TaS$_2$ has a hexagonal structure similar to the parent compound TaS$_2$ with P6$_3$/mmc (194) space group. The obtained lattice parameters are $a = b = 3.315(6)~\text{\AA}$, $c = 12.093(3)~\text{\AA}$ where a slight increase in $c$ parameter is observed as compared to 2H-TaS$_2$ \cite{TaS2_parameter}. However, no significant changes in the lattice parameters indicate that Li atoms have intercalated between the weakly bonded layers instead of residing at atomic sites.
     
\subsection{Anisotropic superconductivity}

\begin{figure*}
\includegraphics[width=1.0\textwidth]{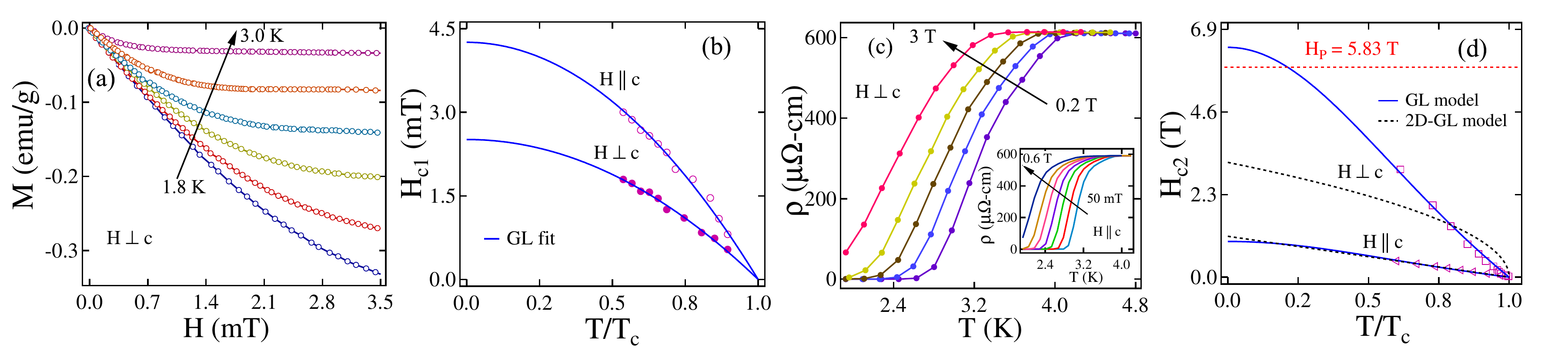}
\caption {\label{Fig2}(a) Field-dependent magnetization plot to calculate lower critical field value. (b) Temperature dependence of the lower critical field values in both directions. The solid blue line shows the GL-fitted curve. (c) Low-temperature resistivity at different fields in $H \perp c$ direction where broad transition indicates effective vortex dynamics in the superconducting region. Inset shows the temperature dependence of resistivity in the $H \parallel c$ direction. (d) Temperature dependence of the upper critical field in both directions manifests anisotropy in the superconductivity. The blue and black lines are fitted to the GL model and 2D-GL model, respectively.  The red dotted line indicates the Pauli limiting field. }
\end{figure*}

\figref{Fig1}(e) presents the zero field resistivity variation with temperature in the $H \parallel c$ direction. As the resistivity decreased with decreasing temperature from 300 to 10 K, it indicated metallic conduction in the normal state. The residual resistivity ratio [RRR = $\rho$(300 K)/$\rho$(6 K)] is obtained as 4.6. In the zero field, an abrupt drop in resistivity is observed at $T_{c}$ = 3.3(1) K, indicating the occurrence of superconductivity in this compound. The inset of \figref{Fig1}(e) (upside) shows the enlarged view of resistivity around the superconducting region. A similar superconducting transition has been reported earlier in Li$_x$TaS$_2$ with $x$ = 0.096 \cite{LixTaS2superconductivity}. As in the normal state resistivity, no change of curvature or any kink was spotted, which is attributed to the suppression of CDW transition by Li intercalation, whereas in parent 2H-TaS$_2$ a CDW anomaly has been reported at $T_{\text{CDW}} \sim$ 75 K \cite{cdw_TaS2}. Normal state resistivity is well fitted by the parallel resistor model, according to which temperature-dependent resistivity is given by -
\begin{equation}
  \rho(T)=\left[\frac{1}{\rho_{sat}}+\frac{1}{\rho_{ideal}(T)}\right]^{-1}
\label{eqn1}
\end{equation}
where $\rho_{sat}$ is the saturated resistivity at high temperature, which is temperature independent, and $\rho_{ideal}(T)$ = $\rho_{i,0} + \rho_{i,L}(T)$  is the temperature independent residual resistivity arising from impurity scattering. $\rho_{i,L}(T)$ can be defined by Wilson's theory as \cite{wilson} -
\begin{equation}
    \rho_{i,L}(T)= C\left(\frac{T}{\theta_{D}}\right)^r\int_{0}^{\theta_{D}/{T}}\frac{x^{r}}{(e^{x}-1)(1-e^{-x})}\,dx
\label{eqn2}
\end{equation}

Here $\theta_{D}$ denotes Debye temperature, and $C$ is the material-dependent constant. After fitting with r = 3, the values of the residual resistivity $\rho_{0}$ = 0.611(2) m$\ohm$-cm and  the Debye temperature $\theta_{D}$ = 165(2) K are extracted.

To calculate the carrier density of Li$_x$TaS$_2$, Hall measurement was performed. The Hall resistivity $\rho_{xy}$ follows a general linear behaviour with varying magnetic field, and the slope value gives the Hall coefficient ($R_H$) as shown in the inset of \figref{Fig1}(e). By linear fit, we obtained $R_H$ = 4.59 $\times$ 10$^{-12}$ cm$^3$/C and accordingly the carrier density $n$ = 1.36 $\times$ 10$^{22}$ cm$^{-3}$ using the formula- $R_H = 1/ne$. The negative slope indicates that electrons are the dominant charge carriers in the system. 

To determine the superconducting transition temperature, magnetic susceptibility was measured under an applied field of 1 mT. \figref{Fig1}(f) displays the temperature dependence of the magnetic susceptibility in both the zero field-cooled warming (ZFCW) and field-cooled cooling (FCC) modes. Below the critical temperature of $T_c$ = 3.29 (6) K, a significant diamagnetic signal was observed, indicating the onset of the superconducting state. The FCC signal was weaker than the ZFCW signal, attributable to strong flux trapping in the sample. Based on ZFCW data, the superconducting  volume fraction was estimated to be $\sim$ 100 $\%$, which confirms the bulk superconductivity in Li$_x$TaS$_2$.

Magnetization-field (M-H) data were collected at various temperatures to determine the lower critical field value, $H_{c1}$, as presented in \figref{Fig2}(a). $H_{c1}(0)$ was estimated by fitting the data as a function of temperature, using the Ginzburg-Landau (GL) relation given in \equref{eqn3: HC1}. The $H_{c1}(0)$ value was obtained from the fit intercept, as shown in \figref{Fig2}(b) for both directions \cite{GLfit}.
\begin{equation}
    H_{c1}(T)= H_{c1}(0)\left[1-\left(\frac{T}{T_{c}}\right)^2\right]
\label{eqn3: HC1}
\end{equation}
The fitting results indicated that the zero temperature values $H_{c1}^{||}$ (0) (parallel to the $c$-axis) and $H_{c1}^{\perp}$(0) (perpendicular to the $c$ axis) were 4.25 (4) mT and 2.51 (2) mT, respectively.

The upper critical field value $H_{c2}$ was determined by measuring the temperature dependence of the resistivity under various magnetic fields for both orientations as $H \parallel c$ and $H \perp c$ as presented in \figref{Fig2}(c). The values of $H_{c2}$ were obtained by setting $\rho$ = 0.1$\rho_n$, where $\rho_n$ represents the resistivity in the normal state. The resulting $H_{c2}$ values as a function of temperature are shown in \figref{Fig2}(d). The data was fitted using the Ginzburg-Landau (GL) and 2D-GL models. For perfect 2D superconductors, $H_{c2}^{\perp}$ follows a $(1-T/T_c)^{1/2}$ behavior in the $H \perp c$ direction. However, the obtained $H_{c2}$ data in the $H \perp c$ direction deviate from the 2D-GL model and are well described by the GL model. According to the GL model, $H_{c2}$ follows:
\begin{equation}
H_{c2}(T)= H_{c2}(0)\left[ \frac{1-(\frac{T}{T_{c}})^2}{1+(\frac{T}{T_{c}})^2}\right]
\label{eqn4:HC2}
\end{equation}
The $H_{c2}$ data in the $H \parallel c$ direction are consistent with both models. Based on the GL model, the estimated values of $H_{c2}^{||}(0)$ and $H_{c2}^{\perp}(0)$ are 0.99(2) T and 6.40(1) T, respectively. The deviation from the 2D-GL model indicates that the superconducting TaS$_2$ layers are not fully decoupled, suggesting quasi-2D superconductivity in Li$_x$TaS$_2$.

The obtained anisotropic parameter $[\Gamma$ = $H_{c2}^{\perp}(0)/H_{c2}^{||}(0)]$ of 6.5, which is slightly lower than that of the parent 2H-TaS$_2$ \cite{anisotropicTaS2}. Na$_{0.1}$TaS$_2$, Cu$_{0.03}$TaS$_2$, and Ni$_{0.04}$TaS$_2$ also show similar anisotropy \cite{NaxTaS2,Cu0.03TaS2,NixTaS2}, which can be attributed to the anisotropy of the Fermi surface. In 2D superconductors, Cooper pairing can be disrupted primarily by the Pauli paramagnetic mechanism, as the orbital effect is negligible due to the constrained electron dynamics in a plane. The Clogston-Chandrasekhar limit, also known as the Pauli paramagnetic limit, is expressed as $H_{P}=1.86 T_{c}$. For Li$_x$TaS$_2$, the Pauli paramagnetic limit is $H_P=5.83$ T, which is lower than the in-plane upper critical field value ($H_{c2}^{\perp}(0)$) and confirms the breaking of the Pauli limit. Possible explanations for this violation in 2D superconductors include Ising spin-orbit coupling (SOC), Fulde-Ferrell-Larkin-Ovchinnikov (FFLO) states, and spin-orbit scattering \cite{pauli_1,pauli_2,pauli_3}. However, a more detailed theoretical investigation is needed to understand the cause of Pauli limit breaking in Li$_x$TaS$_2$.

For anisotropic superconductors, the relation between the upper critical field and the coherence length ($\xi$) can be expressed by \equref{eqn5:coherence} as:
\begin{equation}
    H_{c2} = \frac{\phi_0}{2\pi\xi_{\perp c}^2}(\cos^2\theta+\epsilon^2\sin^2\theta)^{-1/2}
    \label{eqn5:coherence}
\end{equation}
Here, $\phi_{0}$ is the magnetic flux quanta with a value of 2.07$\times$10$^{-15}$ T m$^{2}$, $\theta$ denotes the angle between the applied field and the unit vector perpendicular to the layers, and $\epsilon$ is the ratio of the two coherence lengths, i.e., $\epsilon$ = $\xi_{\parallel c}$/$\xi_{\perp c}$. We can obtain the expressions for the coherence length along the parallel ($\xi_{\parallel c}$) and perpendicular ($\xi_{\perp c}$) directions to the $c$-axis by reducing \equref{eqn5:coherence} for $\theta = 0\degree$ and 90$\degree$, which are given by $H_{c2}^\parallel(0)$ = $\frac{\phi_0}{2\pi\xi^2_{\perp}}$ and $H_{c2}^\perp(0)$ = $\frac{\phi_0}{2\pi\xi_{\parallel}\xi_{\perp}}$, respectively. The values of $\xi_{\parallel c}$ and $\xi_{\perp c}$ were determined to be 2.82(8) nm and 18.20(8) nm, respectively.

\begin{figure}[t]
 \centering
\includegraphics[width=0.92\columnwidth]{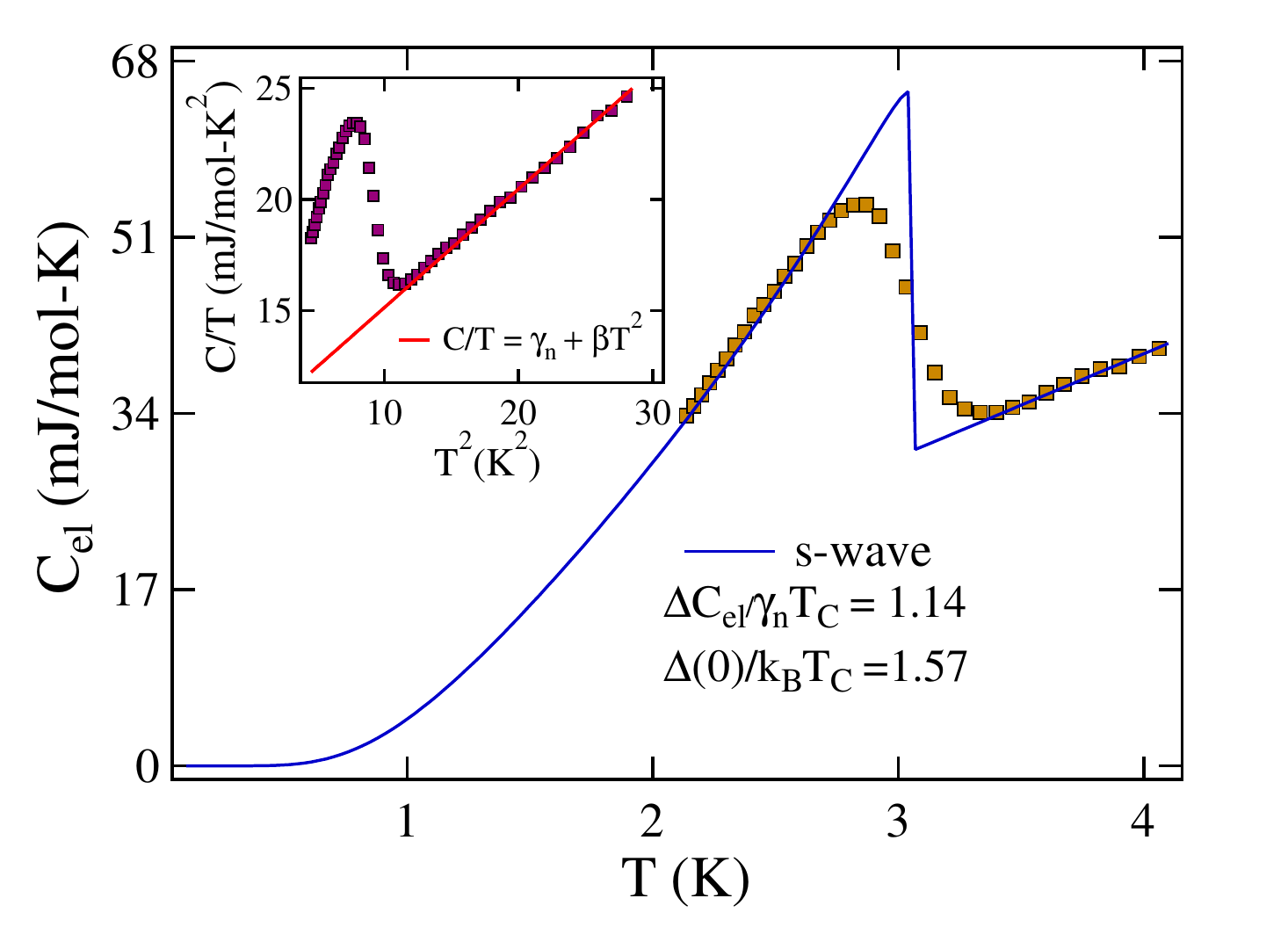}
\caption {\label{Fig3} The electronic contribution to the specific heat is plotted as a function of $T$. The solid blue line displays a BCS-type behaviour. Inset shows the temperature dependence of specific heat where the jump at $T_c\sim$ 3.33(6) K depicts a phase transition from normal to the superconducting state. }
\end{figure}
\begin{table}[h]
\caption{The anisotropic superconducting parameters of Li$_x$TaS$_2$ single crystal }
\label{tbl1}
\setlength{\tabcolsep}{9pt}
\begin{center}
\begin{tabular}[b]{lccc}\hline
Parameters& Unit & $H \parallel c$ & $H \perp c$\\
\hline
$H_{c1}(0)$ & mT & 4.25(4) & 2.51(2) \\ 
$H_{c2}$(0) & T & 0.99(2) &  6.40(3) \\
$\xi$ & nm & 2.82(8) &  18.20(8)  \\
$\lambda_{GL}$ & nm & 838.21(8) & 336.02(2)  \\
$k_{GL}$& & 18.5 & 74.1  \\
\hline
\end{tabular}
\end{center}
\end{table}

The GL penetration length and the GL parameter $\kappa$ were calculated using the equations $H_{c2}^{\perp}(0)/H_{c1}^{\perp}(0) = 2\kappa^2_{\perp c}/\ln{\kappa_{\perp c}}$, $\kappa_{\perp c} = [ \lambda_{\perp c}(0) \lambda_{\parallel c}(0)/\xi_{\perp c}(0)\xi_{\parallel c}(0)]^{1/2}$, and $\kappa_{\parallel c} = \lambda_{\perp c}(0)/\xi_{\perp c}(0)$. All the obtained superconducting parameters are summarized in Table \ref{tbl1}. The value of $\kappa$ > 1/$\sqrt{2}$ indicates that Li$_{x}$TaS$_{2}$ is an extreme type-II superconductor. The thermodynamic critical field, which is a measure of condensation energy, was estimated to be approximately 0.06 T using $H_c$(0) = $H_{c1}^\perp$(0)$\sqrt{2}\kappa_{\perp c}$/ln $\kappa_{\perp c}$. 

The anisotropy ratios for different intercalating atoms in the 2H-TaS$_2$ phase are listed in Table \ref{tbl2} to compare the anisotropy of Li$_x$TaS$_2$. Additionally, the mean-free path ($l_e$) was calculated using the Drude model, $l_e = (3\pi^2)^{1/3}\hbar/e^2\rho_0n^{2/3}$, where $\rho_0$ is the residual resistivity and $n$ is the carrier concentration. The calculated value of $l_e$ was 0.366 nm using the values of $\rho_0$ = 0.611(2) m$\Omega$-cm and $n$. Since $l_e/\xi_0$ = 0.0201 << 1, it indicates that Li$_x$TaS$_2$ is a dirty limit superconductor.

\begin{table}
\caption{Comparison of anisotropy ratio of Li$_x$TaS$_2$ with different intercalated 2H-TaS$_2$ compounds. }
\label{tbl2}
\setlength{\tabcolsep}{8pt}
\begin{center}
\begin{tabular}[b]{lcccl}\hline
Compounds & $T_c$ (K) & $H_{c2}^{\parallel}$ (T) & $H_{c2}^{\perp}$ (T) & $\Gamma$ \\
\hline
Li$_x$TaS$_2$ & 3.3 & 0.99 & 6.40 & 6.5 \\
Na$_{0.1}$TaS$_2$ \cite{Na0.1TaS2} & 4.3 & 2.5 & 16 & 6.2\\
Ni$_{0.04}$TaS$_2$ \cite{NixTaS2} & 4.15 & 4.85 & 17.3 & 3.2 \\
Cu$_{0.03}$TaS$_2$ \cite{Cu0.03TaS2} & 4.03 & 1.8 & 9.6 & 5.3 \\
Pb$_{0.03}$TaS$_2$ \cite{Pb0.33TaS2} & 2.8 & 0.4 & 6.84 & 17.1\\
In$_{0.58}$TaS$_2$ \cite{InxTaS2} & 0.69 & 0.028 & 0.34 & 12\\
\hline
\end{tabular}
\end{center}
\end{table}

\subsection{Specific heat measurement}
The measurement of specific heat confirms the presence of superconductivity in Li$_x$TaS$_2$, as evidenced by the discontinuity in the zero-field specific heat at ~ 3.33(6) K, shown in the inset of \figref{Fig3}. Total specific heat is composed of an electron and a lattice fraction, which can be expressed as $\frac{C}{T}$ = $\gamma{_n}$+$\beta T^2$, where $\gamma_n$ is the Sommerfeld coefficient associated with the electronic contribution and $\beta$ is the Debye constant associated with the phonon contribution. The values of $\gamma_{n}$ and $\beta$ were determined to be 9.77(6) mJ/mol-K$^2$ and 0.534(1) mJ/mol-K$^4$, respectively, by fitting the normal state data denoted by the solid red line in the inset of \figref{Fig3}. Using the \equref{eqn6}, the Debye temperature $\theta_D$ was calculated to be 221(1) K.
\begin{equation}
    \theta_D = \left(\frac{12\pi^4RN}{5\beta}\right)^{1/3}
    \label{eqn6}
\end{equation}
where $R$ is the universal gas constant, and $N$ is the number of atoms per formula unit. The deduced $\gamma_{n}$ value is larger than that of the parent 2H-TaS$_2$ compound.The density of states (DOS) at the Fermi level can be calculated using the formula $\gamma_n$ = $\left(\frac{\pi^2k_B^2}{3}\right) D(E_F)$, resulting in a DOS of 4.15(2) states $eV^{-1}fu^{-1}$ for Li$_x$TaS$_2$. By comparing with 2H-TaS$_2$ \cite{HC_TaS2, TaS2_HC_2}, the observed DOS near the Fermi level is found to be higher in Li$_{x}$TaS$_2$, which may be related to the enhancement of the superconducting transition temperature compared to the parent 2H-TaS$_2$.\\

The strength of the electron-phonon interaction, $\lambda_{e-ph}$, can be estimated according to McMillan's theory \cite{mcmillan} using the relation:
\begin{equation}
 \lambda_{e-ph} = \frac{1.04 + \mu^{*}\ln{(\theta_{D}/1.45T_{c})}}{(1 - 0.62\mu^{*})\ln{(\theta_{D}/1.45T_{c}}) - 1.04 }.
 \label{eqn7: lambda }
\end{equation}
where $\mu^{*}$ is taken to be 0.13 for transition metals, indicating a pseudopotential of effective electron repulsion. Thus, $\lambda_{e-ph}$ is estimated to be 0.62, which manifests Li$_x$TaS$_2$ as a weakly coupled superconductor.

To investigate the superconducting ground state, the electronic specific heat contribution ($C_{el}$) is calculated using $C_{el}(T) = C(T) - \beta T^3$. The low-temperature electronic specific heat plot in \figref{Fig3} fits well with the s-wave model, as observed in Cu$_x$TiSe$_2$ \cite{CuxTiSe2_1}. According to this model, the entropy $S$ for a single-gap BCS superconductor can be expressed as:
\begin{equation}
    S = -\frac{6 \gamma_n}{\pi^2}\left(\frac{\Delta(0)}{k_B}\right)\int_{0}^{\infty}{[f\ln{f} + (1 - f) \ln{(1-f)}]}\,dy
    \label{eqn8}
\end{equation}
where $f(\xi)$ is the Fermi function, defined as $f(\xi) = [\exp{(E(\xi)/k_{B}T)} + 1]^{-1}$, $E(\xi) = \sqrt{\xi^2 + \Delta^2(t)}$ is the excitation energy of the quasiparticles measured relative to the Fermi energy, $y = \xi/\Delta(0)$, $t = T/T_c$, and $\Delta(t)=\tanh{[1.82(1.018(1/t) -1))^{0.51}]}$ is the temperature-dependent superconducting energy gap function. The temperature-dependent electronic specific heat can be calculated as $C_{el}=t\frac{dS}{dt}$. From the fit, the ratio of the superconducting gap $\Delta(0)/k_B T_c$ and the specific heat jump $\Delta C_{el}/\gamma_n T_c$ is found to be approximately = 1.57 and 1.14, respectively. Both values are close to the standard BCS value (1.76 and 1.43) in the weak-coupling limit.
\begin{figure}[t]
\centering
\includegraphics[width=\columnwidth]{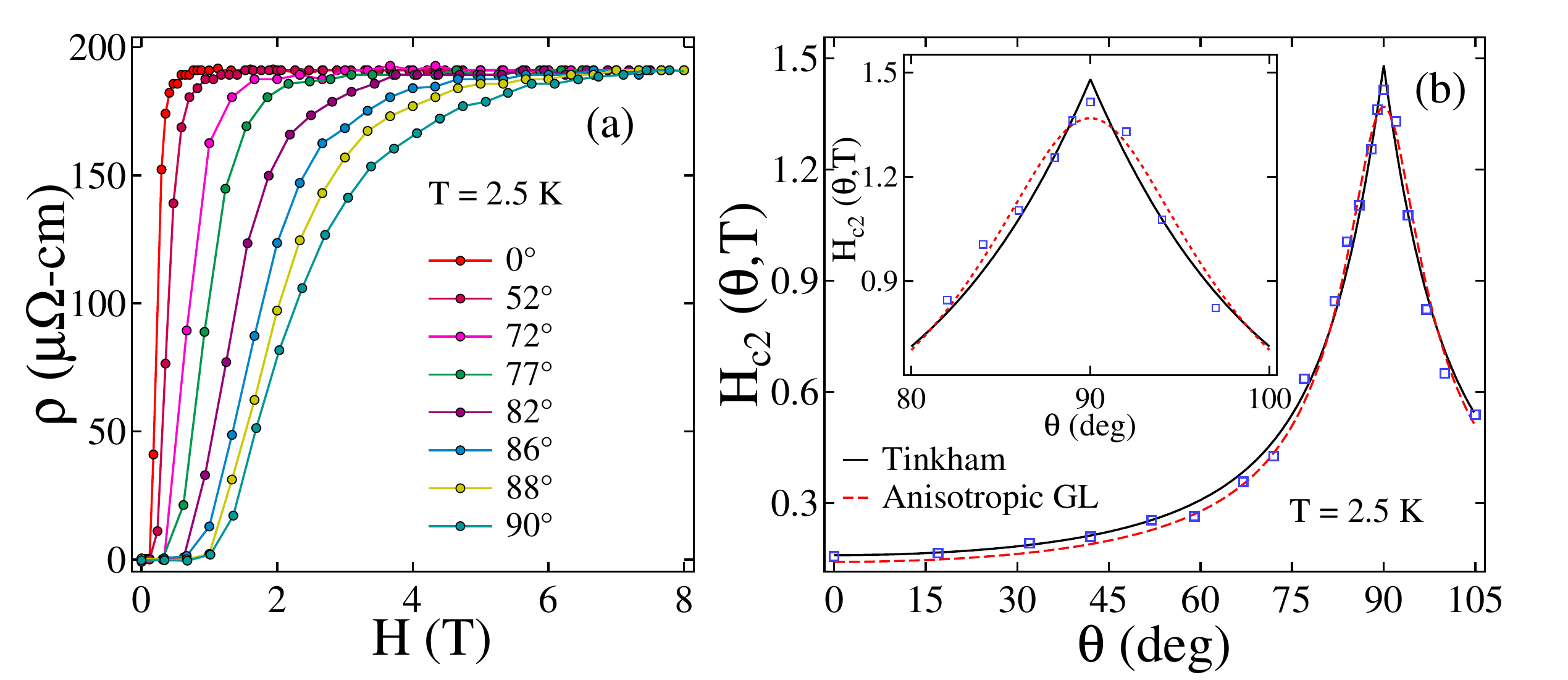}
\caption{\label{fig4}(a) Field dependent resistivity at $T$ = 2.5 K and at different angles $\theta$, where $\theta$ denotes the angle between the $ab$-plane of the crystal and magnetic field direction. (b) Angular dependence of $H_{c2}$ fitted with the 3D anisotropic GL (red) and the 2D Tinkham model (black). In the inset, an enlarged view of $H_{c2}$ ($\theta$) is shown within $\pm$ 10$\degree$ of $\theta$ = 90$\degree$.}
\end{figure}
\subsection{Angle-dependent magnetotransport measurement}
To further explore the 2D superconducting properties, angle-dependent transport measurements of Li$_x$TaS$_2$ crystal. The field-dependent resistivity at different angles at $T$ = 2.5 K is presented in \figref{fig4}(a). The sample was rotated from angle $\theta=0^\circ$ ($H \parallel c$) to $\theta=90^\circ$ ($H \perp c$), with the current always directed along the $ab$-plane. The upper critical field, $H_{c2}$, was calculated at $\rho$ = 0.1$\rho_n$. The angular dependence of $H_{c2}$ is shown in \figref{fig4}(b) with a blue square symbol, which exhibits a cusp-like variation with $\theta$ close to 90$\degree$.

The angular dependence of $H_{c2}$ in anisotropic superconductors is typically explained by two theoretical models. For three-dimensional superconductivity, the angular dependence of $H_{c2}$ can be represented in a simple ellipsoidal form according to the anisotropic Ginzburg-Landau (GL) model \cite{tinkham_intro}:
\begin{equation}
    \left(\frac{H_{c2}(\theta)  \sin{\theta}}{H_{c2}^\perp}\right)^{2} +  \left(\frac{H_{c2}(\theta)  \cos{\theta}}{H_{c2}^\parallel}\right)^{2} = 1
    \label{eqn9:3D}
\end{equation}
Meanwhile, for two-dimensional thin-film superconductors, Tinkham \cite{2dmodel} proposed the following equation: 
\begin{equation}
    \left(\frac{H_{c2}(\theta)  \sin{\theta}}{H_{c2}^\perp}\right)^{2} + \left|\frac{H_{c2}(\theta) \cos{\theta}}{H_{c2}^\parallel}\right| = 1
    \label{eqn10:2D}
\end{equation}

The superconductivity in a 2D regime can be modelled as superconducting layers separated by non-superconducting blocking layers. To determine the nature of superconductivity in Li$_x$TaS$_2$, the extracted $H_{c2}$ values with angles were fitted using both the 2D Tinkham model and the 3D anisotropic GL model shown in \figref{fig4}(b). The data was more consistent with the 2D Tinkham model, indicating a quasi-2D nature of superconductivity in Li$_x$TaS$_2$.

\subsection{Berezinskii-Kosterlitz-Thouless (BKT) transition}

 \begin{figure}
 \centering
\includegraphics[width=1.0\columnwidth]{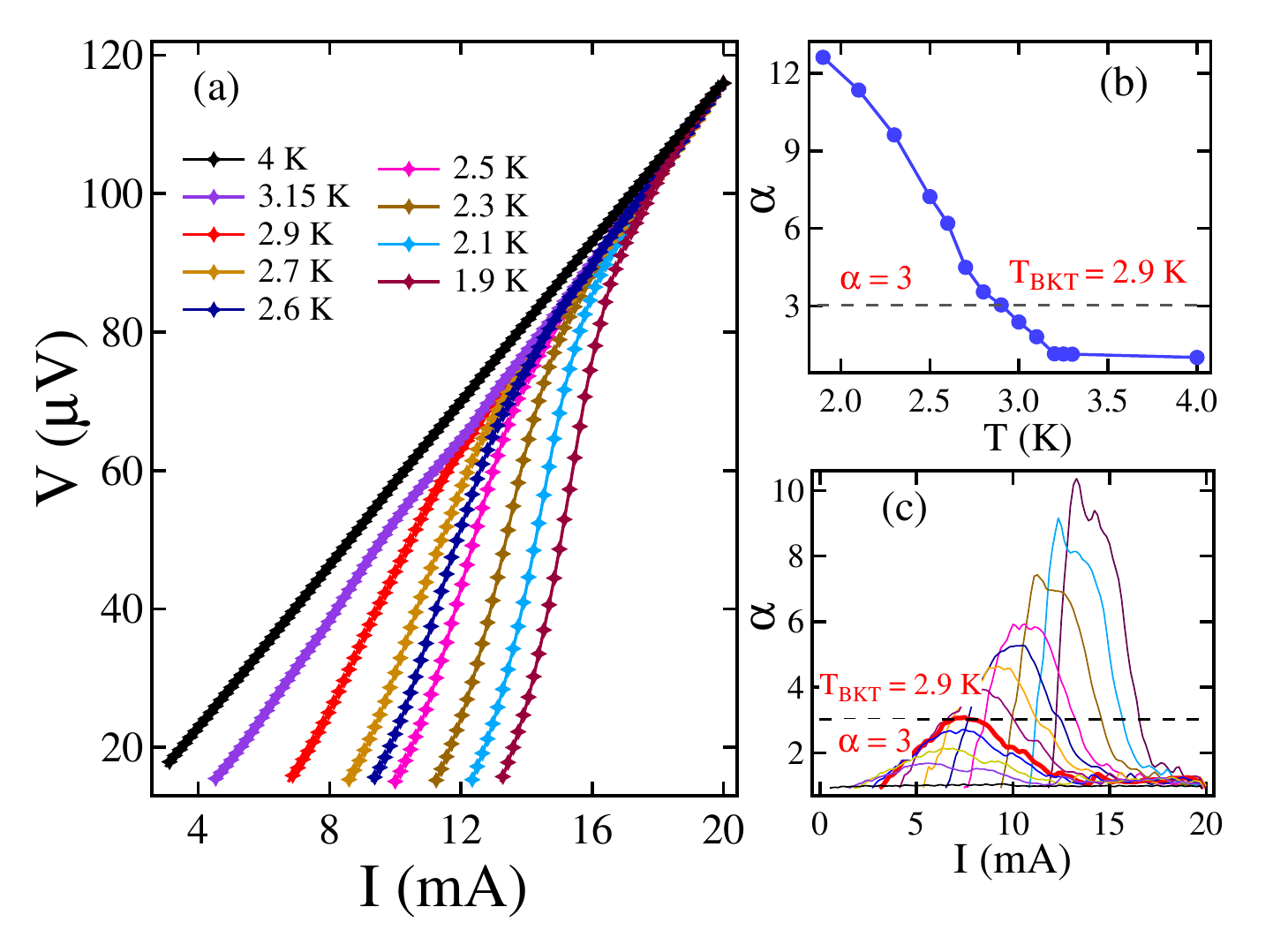}
\caption {\label{Fig5}(a) Temperature dependence of $V-I$ curves at zero applied field to calculate BKT transition temperature. (b) The power-law fitting of the low-current profile calculates the exponent $\alpha$ values. (c) $\alpha$ is also calculated by plotting  $d(\log V)/d(\log I) (I)$ = $\alpha$. The dashed line indicates $T_\text{BKT}$ = 2.9 K corresponds to $\alpha$ = 3. }
\end{figure}

The quasi-2D nature of Li$_x$TaS$_2$ is further supported by the measurement of temperature-dependent $V(I)$ characteristics, which reveal the Berezinskii-Kosterlitz-Thouless (BKT) transition temperature. In systems approaching the 2D limit, strong spatial and temporal fluctuations can destroy phase coherence and superconductivity. However, BKT theory allows the phase transition to occur by establishing quasi-long-range correlations in the order parameter without breaking the symmetry \cite{BKT_1,BKT_2}. While the observation of BKT transition in bulk layered crystals is rare, the $V(I)$ characteristics of Li$_x$TaS$_2$ manifest this transition, as shown in \figref{Fig5}(a). At low temperatures, the $V-I$ characteristics become nonlinear due to the finite critical current and follow a power-law behavior given by-
\begin{equation}
   V \propto I^{\alpha(T)}, ~ \text{with}~ \alpha(T) = 1 + \pi J_s(T)/T
    \label{eqn11:V_I}
\end{equation}
where $J_s$ is the superfluid density. Above $T_c$, the $V-I$ characteristics follow a general linear behaviour corresponding to $\alpha=1$. At the BKT transition, $\alpha$ increases to 3. By fitting the power law in \equref{eqn11:V_I}, we have calculated the exponent values as a function of temperature, as shown in \figref{Fig5}(b). We obtained $T_\text{BKT}=2.9$ K, where $\alpha(T)$ reaches 3. The same $\alpha$ values were also obtained by plotting $d(\log V)/d(\log I)$ as a function of current, as shown in \figref{Fig5}(c).

The quasi-2D behavior observed through the angular dependence of the upper critical field and the BKT transition suggests that intercalating Li atoms in 2H-TaS$_2$ weakens interlayer coupling and significantly changes electronic properties. In the monolayer limit of 2H-TaS$_2$, Ising spin-orbit coupling (SOC) induced robust superconductivity against high magnetic fields \cite{2D-TaS2}, facilitated by the presence of heavy Ta atoms that contribute to strong SOC. Our sample exhibited a high upper critical field that slightly exceeded the Pauli limit in the in-plane direction, prompting interest in investigating the source of the high $H_{c2}$ in the bulk sample. Examining thickness-dependent superconductivity can also lead to a more pronounced 2D nature and enhanced SOC, leading to exotic Ising/FFLO-type superconductivity and other quantum phenomena.
\section{Conclusion}
We conducted transport, magnetization, and specific heat measurements on single crystals of Li-intercalated 2H-TaS$_2$ grown using the chemical vapour transport method. Our results demonstrate that bulk, anisotropic, weakly coupled s-wave superconductivity in Li$_x$TaS$_2$ where the upper critical field ($H_{c2}$) exceeds the Pauli limit. The angular dependence of the upper critical field ($H_{c2}$) conforms to the Tinkham model, denoting quasi-two-dimensional superconductivity. We also observed the Berezinsky-Kosterlitz-Thouless (BKT) transition, affirming the quasi-two-dimensional nature of superconductivity. The high spin-orbit coupling, quasi-2D superconductivity, and upper critical field breaking suggest that the exact superconducting pairing mechanism requires additional low-temperature thickness dependence measurements. Moreover, the readily cleavable nature of 2H-TaS$_2$ provides a unique opportunity to observe quantum phenomena from enhanced spin-orbit coupling, opening doors for further investigations of intercalated 2D materials in the quest for novel quantum behaviour.

\section{Acknowledgments}

T. Agarwal acknowledges the funding agency Department of Science and Technology (DST), Government of India, for providing the JRF fellowship (Award No. DST/INSPIRE/03/2021/002666). R.~P.~S.\ acknowledge Science and Engineering Research Board, Government of India, for the Core Research Grant (CRG/2019/001028).

\end{document}